\documentclass{Interspeech}

\interspeechcameraready
\title{DAFMSVC: One-Shot Singing Voice Conversion with Dual Attention Mechanism and Flow Matching}

\author[affiliation={1,2}]{Wei}{Chen}
\author[affiliation={1}]{Binzhu}{Sha}
\author[affiliation={1}]{Dan}{Luo}
\author[affiliation={2}]{Jing}{Yang}
\author[affiliation={2}]{Zhuo}{Wang}
\author[affiliation={2}]{Fan}{Fan}
\author[affiliation={1,\ast}]{Zhiyong}{Wu}

\affiliation{}{Shenzhen International Graduate School, Tsinghua University}{China}
\affiliation{}{Huawei Technologies Co., Ltd.}{China}
\email{chenw23@mails.tsinghua.edu.cn, zywu@sz.tsinghua.edu.cn}

\usepackage{comment}

\usepackage{graphicx}
\usepackage{amsmath}
\usepackage{multirow}
\usepackage{makecell}
\usepackage{amsmath}
\usepackage{diagbox}
\begin{document}

\maketitle
\renewcommand{\thefootnote}{\fnsymbol{footnote}}

\footnotetext[1]{Corresponding author.}
\renewcommand{\thefootnote}{\arabic{footnote}}

\begin{abstract}
Singing Voice Conversion (SVC) transfers a source singer’s timbre to a target while keeping melody and lyrics.
The key challenge in any-to-any SVC is adapting unseen speaker timbres to source audio without quality degradation.
Existing methods either face timbre leakage or fail to achieve satisfactory timbre similarity and quality in the generated audio.
To address these challenges, we propose DAFMSVC, where the self-supervised learning (SSL) features from the source audio are replaced with the most similar SSL features from the target audio to prevent timbre leakage.
It also incorporates a dual-cross-attention mechanism for the adaptive fusion of speaker embeddings, melody, and linguistic content.
Additionally, we introduce a flow matching module for high-quality audio generation from the fused features.
Experimental results show that DAFMSVC significantly enhances timbre similarity and naturalness, outperforming state-of-the-art methods in both subjective and objective evaluations.
\end{abstract}

\section{Introduction}
In recent years, the application of Singing Voice Conversion (SVC) in music creation has been rapidly emerging.
The goal of any-to-any SVC is to transfer the timbre of a source song to an unseen target singer while preserving the original content and melody.
This technology has a wide range of applications, such as becoming an essential tool for artists and disc jockeys in remixing, sampling, and other creative processes.


The core idea behind any-to-any SVC is to model, disentangle, and utilize various speech attributes, including content, timbre, and pitch.
Previous SVC methods~\cite{li2021ppg,sun2016phonetic} typically rely on pre-trained Automatic Speech Recognition (ASR)~\cite{gulati20_interspeech,yang2023hybridformer} models or SSL based speech models~\cite{chen2022wavlm} and Speaker Verification (SV)~\cite{desplanques20_interspeech,wang23ha_interspeech} to extract content and timbre information from the source and target speech, respectively.
These models then encode the extracted information, which is later decoded to reconstruct the singing audio.
Various deep generative models, such as autoregressive models~\cite{zhang2020durian,takahashi2021hierarchical}, Generative Adversarial Networks (GANs)~\cite{liu2021fastsvc}, Variational Autoencoders (VAEs)~\cite{luo2020singing}, and diffusion models~\cite{liu2021diffsvc}, are used for decoding.
Despite significant advancements in speaker verification technologies, relying solely on the extracted speaker embedding vector to encompass all necessary vocal information remains questionable~\cite{li22da_interspeech}.
Furthermore, effectively separating speaker characteristics from content presents another challenge, often leading to timbre leakage~\cite{chen22g_interspeech}, where some of the source speaker's timbre remains in the converted audio. 
This issue is particularly prominent when using SSL speech models to extract content features~\cite{qian2022contentvec}.

To address the issue of timbre leakage, NeuCoSVC~\cite{sha2024neural} uses an SSL based speech model to extract SSL features from the target speaker’s reference audio and constructs a matching pool.
The SSL features from the source audio are then replaced with the most similar features from the matching pool to achieve timbre conversion.
Since the SSL features used during the conversion phase come directly from the target audio, NeuCoSVC avoids the timbre leakage problem.
However, when replacing the most similar self-supervised features, it overlooks some of the timbre information.
This is because the timbre information is scattered across the entire target set, and only a few features are selected for replacement, resulting in incomplete timbre information and, consequently, a decrease in conversion similarity.
Additionally, the GAN-based waveform reconstruction used in NeuCoSVC faces issues such as instability, mode collapse, and insufficient audio generation quality~\cite{bai2024spa,9746812,liu2024rfwave}.

Therefore, to further improve timbre similarity and audio generation quality, this study introduces a novel any-to-any SVC method with Dual Attention mechanism and Flow Matching (DAFMSVC).
Following NeuCoSVC, DAFMSVC uses a matching pool strategy to prevent timbre leakage.
To enhance the timbre information in the SSL features, we introduce speaker embeddings, which help capture the timbre details scattered across the reference audio.
As pointed out in~\cite{zhou22d_interspeech}, speaker characteristics include not only global timbre information but also local pronunciation variations.
Furthermore, melody, which contains pitch and loudness, is closely tied to content.
Therefore, we introduce a dual cross-attention mechanism module to facilitate the adaptive fusion of speaker embeddings, melody, and linguistic content features.
Additionally, as demonstrated in~\cite{esser2024scaling,le2024voicebox}, flow matching techniques have been shown to provide more stable training and higher sample quality in both image and speech generation.
Building on this, we introduce a conditional flow matching module to improve audio quality, which is trained to predict a vector field and efficiently models the probabilistic distribution of the target audio.



\begin{figure*}[t]
  \centering
  \includegraphics[width=1\linewidth]{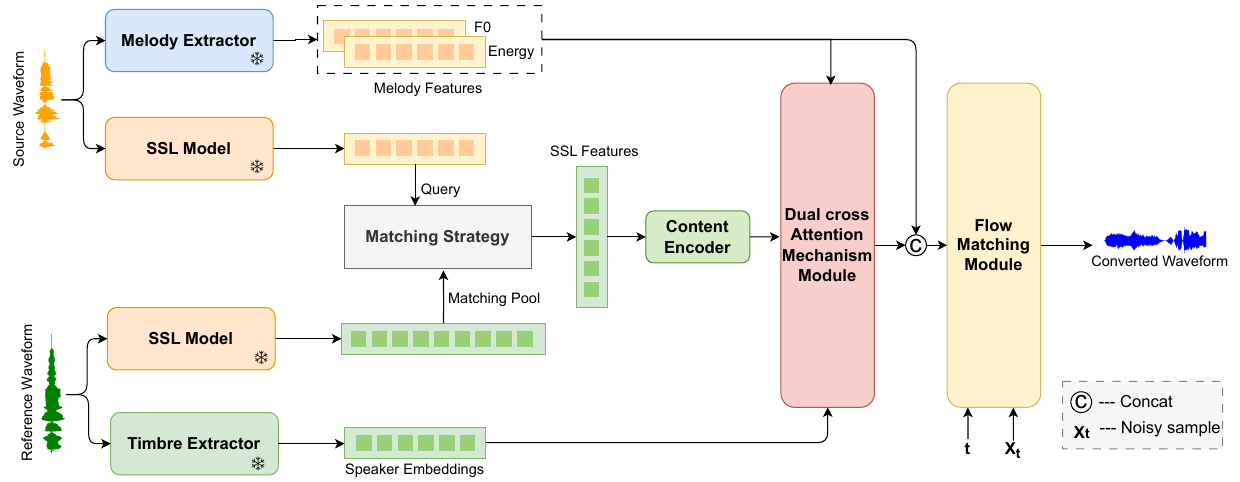}
  \caption{DAFMSVC framework. Snowflake represents the parameter that remains unchanged when training the framework.}
  \label{fig:arc}
\end{figure*}
This work makes the following contributions:
\begin{itemize}
\item We propose DAFMSVC, a novel SVC framework that introduces an innovative dual cross-attention mechanism module with adaptive gate control to effectively capture both timbre and melody information and improve timbre similarity.
\item We introduce a conditional flow matching (CFM) module that predicts probability density paths conditioned on timbre, melody and content, significantly improving sample quality compared to existing state-of-the-art methods.
\item Experimental results show that our model achieves higher timbre similarity and naturalness in both subjective and objective evaluations.
\end{itemize}

\section{Proposed Method}

Figure~\ref{fig:arc} illustrates the overview of our DAFMSVC model.
The source audio is first processed by a pre-trained SSL model to extract fixed-dimensional features that capture both linguistic and timbre information.
These SSL features are then matched with those from the reference audio to select the phonetically relevant ones.
The selected SSL features retain the content information from the source audio while adopting the timbre of the target speaker.
These pre-matched SSL features are encoded and passed through a dual cross-attention mechanism module, enabling the joint utilization of content information, melody, and target timbre representations.
Finally, the output of the dual cross-attention mechanism module is concatenated with pitch and loudness, and fed into the conditional flow matching module to reconstruct the converted waveform.
The details of each module will be discussed in the following sections.
\begin{figure}[t]
  \centering
  \includegraphics[width=1\linewidth]{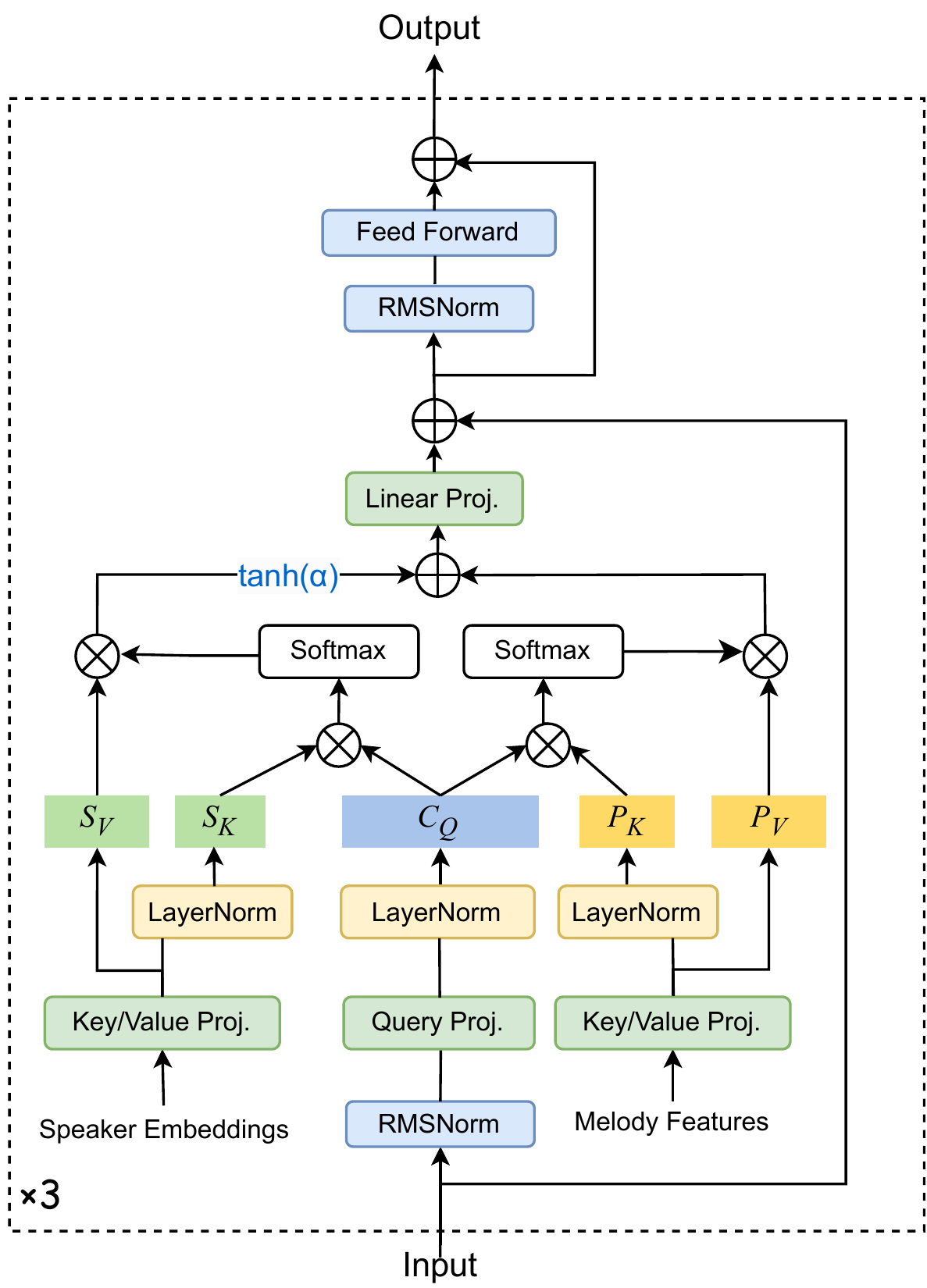}
  \caption{Dual cross-attention mechanism module.}
  \label{fig:c}
  \vspace{-10pt}
\end{figure}

\subsection{Feature extraction and matching module}
The module consists of two stages: 1) extracting compact features from the audio, and 2) replacing the source audio's SSL features with those extracted from the reference utterances.
In the first stage, pitch values are extracted by taking the median of three methods (PYIN\footnote{https://github.com/librosa/librosa}, REAPER\footnote{https://github.com/google/REAPER} and Parselmouth\footnote{https://github.com/YannickJadoul/Parselmouth}).
Loudness features are extracted using the A-weighting mechanism of the power spectrum~\cite{meyer2005measuring}.
These two features are concatenated to form the melody features.
A pre-trained SV model is also used to extract speaker embeddings of the reference waveform.
Next, a pre-trained WavLM-large encoder~\cite{chen2022wavlm} is used to extract SSL features from the audio.
Regarding the matching strategy, following NeuCoSVC~\cite{sha2024neural}, we adopt the KNN method and use the average of the last five layers of WavLM-large to search for the K nearest features in the reference matching pool, as these layers contain more discriminative content information~\cite{lin2023utility}.
The 6th layer of WavLM is then used to replace the features, which are subsequently fed into a content encoder constructed with Feed Forward Transformer blocks~\cite{renfastspeech}.

\subsection{Dual cross-attention mechanism module}\label{sec:a}
To facilitate the adaptive utilization of the content information, melody, and target timbre, we draw inspiration from~\cite{yao2024stablevc} and propose a dual cross-attention mechanism module.
Figure~\ref{fig:c} provides an overview of the attention mechanism module.
Suppose the input of the module is $C$ and the speaker embeddings and melody features denoted as $S$ and $P$, respectively.
${C}_{Q}$ refers to the hidden representation obtained through query projection and query-key normalization.
The goal of timbre attention is to extract fine-grained information from the speaker embeddings.
We use ${C}_{Q}$ as the query for attention, and the embeddings $S$ as both the key and value, allowing the cross-attention mechanism to learn and capture the speaker’s timbre from the speaker embeddings.
To improve the stability of timbre modeling and progressively inject timbre information into both the linguistic content and melody, we introduce an adaptive gating mechanism.
A learnable parameter $\alpha$ with zero initialization is used to control the gating process which ensures stable and consistent modeling of both timbre and melody.
The melody features $P$ serve as both the key and value for melody attention, while ${C}_{Q}$ acts as the attention query.
This structure helps ensure better synchronization between melody variations (such as slides and vibratos) and phoneme boundaries.
The final output $O$ of the dual cross-attention mechanism module is given by the following formula:
\begin{align}
\resizebox{.91\hsize}{!}{$O=\operatorname{softmax}\left(\frac{{{C}_{Q}}{P_{K}}^{T}}{\sqrt{d}}\right) P_{V}+\tanh (\alpha) \operatorname{softmax}\left(\frac{{{C}_{Q}}{{S_{K}}}^{T}}{\sqrt{d}}\right) S_{V}$}
\end{align}
where $d$ is the dimension of queries, $S_{K}$, $S_{V}$ represent the timbre keys and values, while $P_{K}$, $P_{V}$ correspond to the melody keys and values.

Finally, the output of the dual cross-attention mechanism module is concatenated with the melody features, and the final fused features are fed into the CFM module.

\subsection{Conditional flow matching module}
To strike an optimal balance between generation quality and real-time performance, we introduce the conditional flow matching (CFM) module with reference to~\cite{liu2024rfwave}.
Flow matching presents an innovative Ordinary Differential Equation (ODE)-based framework for generative modeling and domain transfer. It introduces a method to learn a mapping that connects two distributions, $\pi_{0}$ and $\pi_{1}$ on $\mathbb{R} ^d$, based on empirical observations:
\begin{align}
\frac{\mathrm{d} Z_{t}}{\mathrm{~d} t}=v\left(Z_{t}, t\right)
\end{align}
where $Z_{0} \sim \pi_{0} \text {, such that } Z_{1} \sim \pi_{1},v:\mathbb{R} ^d \times [0, 1] \rightarrow \mathbb{R}^d$ represents a velocity field.
The training objective is defined as:
\begin{gather}
\resizebox{.91\hsize}{!}{
$L_{rf} = \mathbb{E}_{X_0 \sim \pi_0, (X_1, C) \sim D} \left[ \int_0^1 || (X_1 - X_0)/{\sigma} - \nu(X_t, t | C)/\sigma ||^2 dt \right]$
}
\end{gather}
 where \(\sigma = \sqrt{\text{Var}_1(X_1 - X_0)}\), \(X_t = t X_1 + (1 - t) X_0\) represents a time-differentiable interpolation between $X_0$ and $X_1$ in the time domain, $C$ represents conditional input mentioned in section \ref{sec:a}, $D$ represents the dataset with paired $X_1$ and $C$, and $\text{Var}_1$ calculates the variance along the feature dimension.
Additionally, the model employs multi-band strategies~\cite{yang2021multi} to accelerate audio generation.
To mitigate inconsistencies in subband predictions, Overlap loss $L_{overlap}$ is introduced, while STFT loss $L_{stft}$ is used to reduce artifacts in the presence of background noise.
The overall training loss of DAFMSVC is:
\begin{align}
L = L_{rf} + \lambda\times(L_{overlap} + L_{stft})
\end{align}
where $\lambda$ is set to 0.01, following the parameter settings in~\cite{liu2024rfwave}.

Finally, We sample from the standard Gaussian distribution as the initial condition at $t=0$. By using 10 Euler steps, we approximate the solution to the ODE, effectively generating samples that match the target distribution.



%

%


\section{Experimental Setup}

\subsection{Dataset}
Experiments are conducted on the OpenSinger dataset~\cite{huang2021multi}, which is recorded in a professional studio and contains 50 hours of high-quality Chinese singing.
This dataset includes 28 male singers and 48 female singers, with the audio saved in wav format at a sampling rate of 44.1 kHz.
The singing of two male and two female singers is reserved for the test set, while the remaining recordings are randomly split into the training and validation sets with a 9:1 ratio.

\subsection{Training conditions}
Pitch and loudness features are extracted from 24kHz audio with a hop size of 240.
Notably, to adapt the pitch to the target speaker’s vocal range, the source pitch values are scaled by a shift factor during conversion.
This factor is the ratio of the median pitch in the target audio to that in the source audio.
We utilize a pre-trained state-of-the-art SV model\footnote{https://www.modelscope.cn/models/iic/speech\_campplus\_sv\_zh-cn\_16k-common/summary}
, called CAM++~\cite{wang23ha_interspeech}, to extract speaker embeddings.
This model is trained on a large Chinese speaker dataset, which includes approximately 200k speakers.
A pre-trained WavLM-Large~\cite{chen2022wavlm} is used to extract 1024-dimensional SSL features.
In the matching strategy, the k-nearest method is employed with $ k=4$, and cosine similarity is used as the distance metric, following~\cite{sha2024neural}.
The CFM module, which consists of a ConvNeXtV2~\cite{woo2023convnext} backbone, takes the fused features as input and generates 24kHz singing audio.
The training setup strictly follows~\cite{liu2024rfwave}, except for the number of channels in the conditional input, which is 258.
The AdamW optimizer with an initial learning rate of 0.002 is used for training.
During the inference stage, we sample the waveform using 10 Euler steps within the CFM module, with a guidance scale of 1.0 applied.

\subsection{Baselines}
We evaluate the one-shot SVC performance of DAFMSVC by comparing it with three state-of-the-art systems: NeuCoSVC, DDSP-SVC, and So-VITS-SVC.
NeuCoSVC\footnote{https://github.com/thuhcsi/NeuCoSVC} is a novel neural concatenation-based approach for one-shot SVC, which adopts the FastSVC architecture to generate synthesized audio.
DDSP-SVC\footnote{https://github.com/yxlllc/DDSP-SVC} is an end-to-end singing voice conversion system based on Differentiable Digital Signal Processing (DDSP) that uses a cascade diffusion model to reconstruct high-quality audio.
So-VITS-SVC\footnote{https://github.com/svc-develop-team/so-vits-svc} is a popular open-source voice conversion tool based on VITS~\cite{kim2021conditional}, which uses a Conditional Variational Autoencoder combined with Adversarial Learning.
To ensure a fair comparison, all methods are trained on the same dataset.
Audio samples can be found in demo pages\footnote{https://wei-chan2022.github.io/DAFMSVC/}.

\begin{table*}[th]
    \caption{Comparison with state-of-the-art methods}
    \label{tab:compare}
    \centering
    \scalebox{1}{
    \begin{tabular}{l c c c c c c} \toprule
         Method&  F0CORR$\uparrow$& Loudness RMSE$\downarrow$  & SSIM$\uparrow$  &  MCD$\downarrow$ &  MOS-Naturalness$\uparrow$  & MOS-Similarity$\uparrow$ \\ \midrule
Source& - &   -  & -  & -   &  4.69±0.07  & -\\
DDSP-SVC & 0.909 & 0.129 &  0.600 &8.941 &  2.02±0.09& 2.07±0.08\\ 
So-VITS-SVC& 0.946& 0.155&  0.602&8.227 & 3.45±0.10& 3.06±0.10\\ 
NeuCoSVC& 0.942& 0.114& 0.692&8.634& 3.47±0.11&3.48±0.11\\ 
DAFMSVC&\textbf{0.948}& \textbf{0.067}& \textbf{0.754}&\textbf{7.220}& \textbf{3.80±0.09}&\textbf{3.58±0.11}\\
  \bottomrule
 \multicolumn{6}{l}{MOS results are reported with 95\% confidence intervals.}\\
    \end{tabular}
    }
    \vspace{-8pt}
\end{table*}


\subsection{Evaluation metrics}
We conduct both objective and subjective evaluations to assess the model performance.

For the objective evaluation, we use cosine distance of extracted speaker embeddings (singer similarity, SSIM) to assess singer similarity, F0CORR and Loudness RMSE to measure the naturalness of the converted waveforms~\cite{huang2023singing}, and Mel Cepstral Distortion (MCD) to evaluate audio quality.
1) SSIM: we use the pre-trained CAM++ speaker verification model~\cite{wang23ha_interspeech} to evaluate the singer similarity between the generated samples and the target speaker reference.
2) F0CORR: we evaluate pitch accuracy by calculating the Pearson correlation coefficient of the F0 contours between the source and converted audio, using dynamic time warping (DTW) to align the sequences before comparison. Note that the F0 sequence is normalized using min-max scaling before processing.
3) Loudness RMSE: we compute the root mean square error (RMSE) in loudness between the converted waveform and the source waveform, as the source waveform contains the real speaker’s fine-grained prosody and naturalness.
3) MCD: we adopt the Pysptk tools to extract the Mel-frequency cepstral coefficients (MFCCs) and use DTW to align the target and converted audio parameters.
A lower value indicates higher similarity.

For subjective evaluation, we conduct a Mean Opinion Score (MOS) test with a 5-point scale (1 - bad, 2 - poor, 3 - fair, 4 - good, 5 - excellent).
We invite 15 volunteers with extensive knowledge of music theory to assess the similarity and naturalness of the audio.


\section{Experimental Results}

\subsection{Comparison with state-of-the-art methods}
In the objective experiments, we randomly select 37 audio samples from the validation set and convert them to four unseen target speakers in the test set, resulting in a total of 148 samples.
Table \ref{tab:compare} presents the results of objective evaluations.
Our model outperforms the baseline systems across all metrics, especially in singer similarity.
NeuCoSVC effectively prevents timbre leakage through a SSL features replacement strategy, improving timbre similarity.
Building upon this, DAFMSVC uses a dual cross-attention mechanism module to capture fine-grained timbre details from the speaker embeddings vectors and melody information, leading to a significant enhancement in singer similarity and naturalness.
In terms of MCD, DAFMSVC also achieves a lower score, indicating higher generation quality, owing to the excellent generation capability of the CFM module.
In comparison, NeuCoSVC's audio quality is slightly inferior to that of So-VITS-SVC, possibly due to its GAN-based FastSVC architecture for audio generation~\cite{liu2024rfwave}.

In the subjective experiments, we select 20 audio samples for testing, with five converted samples for each unseen speaker.
As shown in Table \ref{tab:compare}, the results demonstrate that, compared to methods that use speaker embeddings or SSL features replacement for timbre conversion, DAFMSVC achieves better similarity and naturalness.


\subsection{Ablation study}
In this subsection, we conduct ablation studies to evaluate the contribution of each component to timbre modeling and naturalness. Specifically, we perform the following experiments:
1) without speaker embeddings and the dual cross-attention mechanism module, where only the encoded content and melody features are concatenated and passed into the CFM module, referred to as ``\textbf{-spk\&att}'';
2) using speaker embeddings but without the dual cross-attention mechanism module, where the individual features are simply concatenated and passed into the dual cross-attention mechanism module, referred to as ``\textbf{-att}''.

As shown in Table \ref{tab:ablation}, when the speaker embeddings input is removed, the timbre similarity significantly decreases.
This occurs because, during the SSL features replacement, only a small amount of scattered timbre information is incorporated into the SSL features.
As a result, the model is unable to fully capture the target speaker's characteristics, leading to a decrease in timbre similarity.
Notably, although the timbre similarity is lower than that of DAFMSVC, the performance still surpasses NeuCoSVC (as in Table \ref{tab:compare}), due to the powerful generative capability of the CFM module.
When the dual cross-attention mechanism module is removed, both SSIM and MCD metrics degrade, indicating that the simple concatenation of features hinders the CFM from learning how to reconstruct the audio effectively.
In contrast, the attention mechanism helps the model more effectively capture the intricate relationships between timbre, melody, and content, leading to more precise and coherent audio generation.

\begin{table}[t]
  \caption{The ablation study results}
  
  \label{tab:ablation}
  \centering
  \begin{tabular}{>{\arraybackslash}p{0.15\linewidth}>{\centering\arraybackslash}p{0.15\linewidth}>{\centering\arraybackslash}p{0.29\linewidth}>{\centering\arraybackslash}p{0.08\linewidth}>{\centering\arraybackslash}p{0.08\linewidth}}
    \toprule
    Model &   F0CORR$\uparrow$& Loudness RMSE$\downarrow$ & SSIM$\uparrow$ &  MCD$\downarrow$  \\
    \midrule
    DAFMSVC       & 0.948 & 0.067& 0.754& 7.220        \\
    - spk\&att    & 0.947 & 0.109& 0.709& 7.888        \\
    - att         & 0.945 & 0.103& 0.710& 8.129        \\
    \bottomrule
  \end{tabular}
\end{table}

\section{Conclusion}
In this paper, we presents DAFMSVC, a novel any-to-any SVC framework that enhances timbre similarity and improve audio quality.
By combining SSL features with a matching pool strategy, DAFMSVC effectively prevents timbre leakage. 
It also employs a dual cross-attention mechanism module to adaptively fuse speaker embeddings, pitch, and linguistic content features, generating high-quality fused representations that address the timbre similarity issues caused by SSL features replacement.
Additionally, the introduction of the flow matching module significantly improves the reconstruction of high-quality audio.
Experimental results show that DAFMSVC achieves superior timbre similarity and naturalness in both subjective and objective evaluations, outperforming existing state-of-the-art methods.
Future work will focus on further improving the model's efficiency and exploring the application of our proposed method in complex noisy environments.

\section{Acknowledgements}

This work is supported by National Natural Science Foundation of China (62076144) and Shenzhen Science and Technology Program (JCYJ20220818101014030).

\bibliographystyle{IEEEtran}
\bibliography{mybib}
\vspace{-10pt}

\end{document}